\documentclass[]{aa}
\usepackage{epsfig}

\newcommand{\Rsun}{\mbox{R$_{\scriptsize \odot}$}} 
\newcommand{\etal}{{\it{et al.~}}}
\newcommand{\eg}{{\it{e. g.~}}}
\newcommand{\ie}{{\it{i. e.~}}}

\newcommand{\degrees}{$^{\circ}$}

\newcommand{\HII}{\mbox{H\,{\sc ii}}}

\def\mum{\hbox{\,$\mu$m}}

\def\cm{\hbox{\,cm}}

\def\cm_2{\hbox{\,cm$^{-2}$}}
\def\cm_1{\hbox{\,cm$^{-1}$}}

\def\megjysr{\hbox{\,MJy/sr}}

\begin{document}

\title { The Aromatic Infrared Bands as seen by ISO-SWS: 
	probing the PAH model~ 
        \thanks{Based on observations with ISO, an ESA project 
        with instruments funded by ESA Member States (especially the PI 
        countries: France, Germany, the Netherlands and the United Kingdom) 
        and with the participation of ISAS and NASA.} }

\author{L. Verstraete\inst{1}, C. Pech\inst{2}, C. Moutou\inst{3}, K. Sellgren\inst{4}, C.M. Wright\inst{5}, 
        M. Giard\inst{2}, A. L\'eger\inst{1}, R. Timmermann\inst{6}, S. Drapatz\inst{6} }

\institute{$^1$Institut d'Astrophysique Spatiale, B\^{a}t. 121,
           Universit\'e de Paris XI, 91405 Orsay, France \\
           $^2$CESR, 9 Avenue du Colonel Roche, 31028 Toulouse Cedex, France \\
           $^3$Alonso de Cordoba 3107, Santiago 19, Chile \\
           $^4$Astron. Dept., Ohio State University, 140 West 18th Avenue, 
           Colombus OH 43210, USA \\
           $^5$Univ. College, ADFA, UNSW Canberra, Australia \\
           $^6$Max-Planck Inst. f\"ur extraterr. Physik, Postfach 1603, D-85740 
               Garching, Germany
           }

\offprints{L. Verstraete (Laurent.Verstraete@ias.u-psud.fr)}

\date{Received October 2, 2000; accepted April 3, 2001}

\abstract{ We discuss the Aromatic Infrared Band (AIB) profiles observed by ISO-SWS towards
a number of bright interstellar regions where dense molecular gas is 
illuminated by stellar radiation. Our sample spans a broad range of excitation 
conditions (exciting radiation fields with effective temperature, $T_{eff}$, ranging from 23,000 to 
45,000 K). The SWS spectra are decomposed coherently in our sample
into Lorentz profiles and a broadband continuum. We find that the individual 
profiles of the main AIBs at 3.3, 6.2, 8.6 and 11.3 \mum~ are well represented with at most two lorentzians.
The 7.7 \mum-AIB has a more complex shape and requires at least three Lorentz profiles.
Furthermore, we show that the positions and widths of these AIBs are remarkably stable (within a few \cm_1) 
confirming, at higher spectral resolution, the results from ISOCAM-CVF and ISOPHOT-S. This spectral 
decomposition with a small number of Lorentz profiles implicitly assumes that most of the observed bandwidth arises from a 
few, large carriers. Boulanger \etal (1998$b$) recently proposed that the AIBs are the intrinsic profiles of resonances 
in small carbon clusters. This interpretation can be tested by comparing the AIB profile parameters 
(band position and width) given in this work to laboratory data on relevant species when it becomes available.\\ 
Taking advantage of our decomposition, we extract the profiles of individual AIBs from the data and compare 
them to a state-of-the-art model of Polycyclic Aromatic Hydrocarbon (PAH) cation emission. In this model, the 
position and width of the AIBs are rather explained by a redshift and a broadening of the PAH vibrational bands 
as the temperature of the molecule increases (Joblin \etal 1995). In this context, the present similarity of the AIB profiles 
requires that the PAH temperature distribution remains roughly the same whatever the radiation field hardness. 
Deriving the temperature distribution of interstellar PAHs, we show that 
its hot tail, which controls the AIB spectrum, sensitively depends on $N_{min}$ 
(the number of C-atoms in the smallest PAH) and $T_{eff}$. Comparing the observed profiles
of the individual AIBs to our model results, we can match all the AIB profiles (except the 
8.6 \mum-AIB profile) if $N_{min}$ is increased with $T_{eff}$. This increase is naturally explained 
in a picture where small PAHs are more efficiently photodissociated in harsher radiation fields.
The observed 8.6 \mum-profile, both intensity and width, is not explained by our model.\\ 
We then discuss our results in the broader context of ISO observations
of fainter interstellar regions where PAHs are expected to be in neutral form.
}

\authorrunning{L. Verstraete \etal}
\titlerunning{The AIB profiles in ISO-SWS}
\maketitle

\section{Introduction}
The family of infrared features at 3.3, 6.2, 7.7, 8.6, 11.3 and 12.7 \mum~ has been
observed towards a large number of sightlines in the Galaxy and in other
galaxies since the nineteen seventies. Early on, it was recognized that these bands 
correspond to vibrational modes in carbonaceous aromatic systems (Duley \& Williams 1981,
L\'eger \& Puget 1984, Allamandola \etal 1985). These dust bands are therefore dubbed 
the {\em Aromatic Infrared Bands} (hereafter AIBs). The bands at 3.3, 8.6, 11.3 and
12.7 \mum~
stem from vibrational modes of the aromatic C-H bond; the remaining bands are ascribed to
vibrations of the aromatic C-C bonds. The Infrared Space Observatory 
(ISO, Kessler \etal 1996) mission has provided us with an unprecedented wealth of
data in this context.\\

Comparative studies of the AIBs in a wide variety of environments have been made possible 
by the high sensitivity of the ISO camera (ISOCAM, Cesarsky C. \etal 1996) with its circular 
variable filter and of the ISOPHOT-S spectrophotometer (Lemke \etal 1996), which both have low
spectral resolutions ($\lambda / \Delta\lambda \sim 40$ and 90 respectively: 
Boulanger \etal 1996, 1998$a$ and 1998$b$; Cesarsky \etal 1996$a$, 1996$b$, 2000$a$ and 2000$b$;
Cr\'et\'e \etal 1999; Klein \etal 1999; Laureijs \etal 1996; Mattila \etal 1996; Persi \etal 1999; 
Uchida \etal 1998 and 2000). 
The AIB profiles as seen in these data are very similar (in position and width) over 
a range of objects where the stellar radiation field and effective temperature vary 
greatly (1 to 10$^4$ times the standard interstellar radiation field, $T_{eff}$=11,000 
to 50,000 K). The profile invariance as well as the large width of the AIBs lead Boulanger 
\etal (1998$b$) to conclude that the carriers of these bands are large aromatic systems 
containing more than 50 C-atoms. \\

The Short Wavelength Spectrometer (SWS, de Graauw \etal 1996) onboard ISO, less sensitive 
than ISOCAM but with higher spectral resolution ($\lambda/\Delta\lambda$=1000) and broader 
wavelength coverage (2.4-45 \mum), brings a better view of the interstellar AIBs in the
brightest regions. Such detailed data have the ability to constrain the nature and
physical state of the band carriers (Beintema \etal 1996, Molster \etal 1996, Roelfsema \etal 1996, 
Verstraete \etal 1996, Moutou \etal 2000, van Kerckhoven \etal 2000, Hony \etal 2000). In the first part 
of this paper, we decompose the AIB spectrum into Lorentz profiles and a broadband continuum in order 
to characterize (band position and width) each individual AIB and to compare them between objects. 
In the second part of the paper, we confront these 
new observations with a model considering free-flying aromatic molecules (Polycyclic Aromatic Hydrocarbons
or PAHs) as the origin of the AIBs. Indeed, the presence of AIBs in the low-excitation 
diffuse interstellar medium (Boulanger \etal 1996, Mattila \etal 1996, Onaka \etal 1996) requires the 
existence of free-flying PAHs or small grains excited by starlight; furthermore, these PAHs or grains 
must be small enough to undergo strong temperature fluctuations leading to emission of the AIBs in the 
near-infrared (Sellgren 1984). In this emission mechanism, the shape of the emergent AIB spectrum only depends 
on the radiation field hardness (or $T_{eff}$, see Sect. 2) and not on the flux of stellar photons (parameterized 
as $\chi$ in Sect. 2).

In Sect. 2, we present the SWS spectra. The observed AIB profiles are characterized in Sect. 3. 
These results are compared to the predictions of the PAH
model in Sect. 4. We summarize and discuss the significance of our results
in Sect. 5.

\section{Observations and physical conditions}

SWS has observed the AIB spectrum along a number of interstellar sightlines
covering a wide range of excitation conditions. We discuss here three
spectra which sample well the radiation field sequence covered by
the SWS observations. The lines of sight selected all correspond to {\em interface} conditions, \ie,  
regions where fresh molecular material is directly exposed to the stellar light. At interfaces, 
the AIB emission is usually strong (probably because of an enhanced PAH abundance, Bernard \etal 1993):
this is why we selected such regions to carry out the present study.

At the low excitation end, we have the reflection 
nebula NGC 2023 (TDT=65602309, SWS01 speed 3). This spectrum has already been presented in Moutou \etal (1999).
SWS looked at a filament, 60$''$ south of the central star (central star: RA= 5$^{\rm h}$ 41$^{\rm m}$ 38.3$^{\rm s}$, 
DEC= $-$2\degrees 16$'$ 32.6$''$), which is bright in fluorescent H$_2$ emission (Field \etal, 1998).
At the high excitation end, we present here a spectrum of the M17-SW photodissociation interface 
(RA= 18$^{\rm h}$ 20$^{\rm m}$ 22.1$^{\rm s}$, DEC= $-$16\degrees 12$'$ 41.3$''$; TDT=32900866, SWS01 speed 4): this is position 
number 6 of the data presented in Verstraete \etal (1996). Finally, we also have the Orion Bar 
(RA= 5$^{\rm h}$ 35$^{\rm m}$ 20.3$^{\rm s}$, DEC= $-$5\degrees 25$'$ 20$''$; TDT=69501806, SWS01 speed 4) at the position of the 
peak of fluorescent H$_2$ emission (van der Werf \etal 1996). All positions given above are in J2000.

The data reduction was undertaken with the SWS-IA3 environment running at the Institut 
d'Astrophysique Spatiale, Orsay. The spectrum of NGC 2023 lacks a small range around 4 \mum~ because
of bad dark current measurements.
The flux calibration files CAL-G version 030 have been used. 
For the beam sizes, we took the values recently determined by Salama (2000).
This assumes that the source completely fills the beam. To check this assumption, we compared 
our SWS spectra of M17-SW and of the Orion Bar to CAM-CVF data (Cesarsky D. \etal 1996 and 
Cesarsky D. \etal 2000$b$): the continuum fluxes (per solid angle) of the two instruments were found to agree 
within 20\%. In the case of NGC 2023, the emission seen in the ISOCAM-map of Abergel \etal (2000, 
in preparation) looks homogeneous at the position of our SWS spectrum. 
The 6 arcsecond pixels of ISOCAM are much smaller than the SWS field of view, and since the ISOCAM 
image of each of our sources is smooth in the region observed by SWS, we can safely say
that our sources uniformly fill the SWS beam. This statement only holds over
the 5-16 \mum~ wavelength range. In fact, to assure continuity in our spectra, we had to deviate from the 
Salama SWS beam sizes above 27 \mum~ (the spectral bands 3E and 4 of the SWS, see de Graauw \etal 1996).

\begin{figure}[!h]
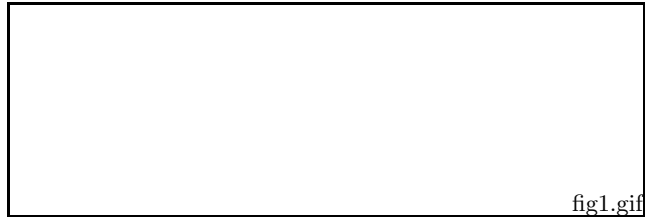

\framebox(240,80)[br]{fig1.gif}
 \caption{ The SWS 2.4-25 \mum~ spectra of the AIBs towards various interstellar sources (see text).
           The spectra have been normalized to the 6.2 \mum-feature of NGC 2023; the scaling factors 
	   are given in the figure, {\it e.g.}, M17-SW/3 means that the flux of M17-SW has been divided by 3.
	   A positive constant has been added to all spectra except that of NGC 2023 to ease comparison.
	   The zero flux level of each spectrum is indicated by the dashed line. The resolving power 
           ($\lambda/\Delta\lambda$) is 200 for NGC 2023 and 500 for M17-SW and the Orion Bar. 
           The narrow, unresolved lines are due to Br$\beta$ 2.63\mum~, Br$\alpha$ 4.06\mum~, 
           forbidden ionic lines ([ArII] 6.98\mum, [ArIII] 8.99\mum, 
           [SIV] 10.5\mum, [NeII] 12.8\mum, [NeIII] 15.5\mum~ and [SIII] 18.7\mum) and H$_2$ pure rotational 
           lines (0--0 S(5) 6.91\mum, 0--0 S(4) 8.02\mum, 0--0 S(3) 9.66\mum, 0--0 S(2) 12.3\mum~ and 0--0 S(1) 17.0\mum).
	  }
\end{figure}

The resulting 2.4-25 $\mu$m spectra are shown in Fig. 1. As we will see below, this reduced spectral 
range amply suffices in our discussion of the AIB profiles. The NGC 2023 spectrum has a resolving power
$\lambda/\Delta\lambda$ = 200 and those of M17-SW and of the Orion Bar have $\lambda/\Delta\lambda$ = 500. 
The physical conditions prevailing in each interstellar region are summarized
in Table 1. We assumed that the radiation field from the exciting star is 
a blackbody characterized by an effective temperature. In this table, we give the effective 
temperature of the exciting
star ($T_{eff}$), the dilution factor for the blackbody ($W_{dil}$), the flux scaling factor ($\chi$) 
in units of the Habing field (at
$\lambda$ = 1000 \AA) and $d_{\star}$ the distance of our line of sight from the central star 
(we assumed $d_{\star}$ to be the projected distance on the sky in all cases). 
In the case of M17-SW, several stars excite the regions we observed: 
in Table 1 we give effective values ($T_{eff}$, $R_{\star}$) that reproduce well the sum 
of all radiation fields.
Our excitation sequence thus goes from NGC 2023 (cool star)
to M17-SW (hot star).

\begin{table}[!ht]
\caption{Physical conditions for the AIB spectra - see text}
\begin{center}
\leavevmode
\begin{tabular}[h]{cccc}
\hline
\multicolumn{4}{c}{} \\
\multicolumn{1}{c}{Line of sight} & \multicolumn{1}{c}{NGC 2023} & 
\multicolumn{1}{c}{Orion Bar} & \multicolumn{1}{c}{M17-SW} \\
\multicolumn{4}{c}{} \\
\hline
\hline
\multicolumn{4}{c}{} \\
$T_{eff}$ (10$^4$ K)     &  2.3$^a$  &  3.7$^c$  &  4.5$^e$  \\
\multicolumn{4}{c}{} \\
$R_{\star}$ (\Rsun)      &  3.2$^{a,b}$  & 8.5$^b$    & 12     \\
\multicolumn{4}{c}{} \\
$d_{\star}$ (pc)         &  0.14   &  0.24 &  1.10  \\
\multicolumn{4}{c}{} \\
$W_{dil}/10^{-13}$  &  2.66 &  6.38 &  0.96 \\
\multicolumn{4}{c}{} \\
$\chi/10^3$           &  1.2  & 42    & 12.5  \\
\multicolumn{4}{c}{} \\
\hline
\end{tabular}
\end{center}
The distances to NGC 2023, M17-SW and the Orion Bar were taken to be 480 pc$^a$, 2.2 kpc$^e$ 
and 460 pc$^d$ respectively.

References: $(a)$ Buss \etal 1994, $(b)$ Lang 1991, $(c)$ Rubin \etal 1991, 
$(d)$ van der Werf \etal 1996, $(e)$ Felli \etal 1984
\end{table}

\section{The observed AIB profiles}

The continuum below the AIBs shows a clear evolution from NGC 2023 to the Orion Bar (see Fig. 1): 
continuous emission of hot, small (radii of a few 10 to 100 \AA, D\'esert \etal 1990) grains between 
10 and 20 \mum~ is prominent in M17-SW and the Orion Bar whereas this is completely absent in NGC 2023. 
This is a consequence of the stronger flux and harder radiation field in the Orion Bar which heats up these 
small grains. 

The 6.2 and 7.7 \mum~ bands do not vary much with respect to each other whereas they do 
relative to the 3.3, 8.6 and 11.3 \mum~ AIBs. Such variations cannot be interpreted
as the result of different emission temperature distributions of PAHs (obtained
with different size distributions and/or radiation field effective temperature). 
In fact, the 3.3 \mum-band is dominated by the emission of small (hot) PAHs while 
the 11.3 \mum-band is contributed to by larger (cold) PAHs (Schutte \etal 1993): thus, changing 
the PAH temperature distribution 
cannot explain why the 11.3 \mum-band varies along with the 3.3 \mum-band. Instead, modifications of the 
physical state of PAHs have to be invoked. Theoretical (de Frees \etal 1993; Pauzat \etal 1995,1997; 
Langhoff 1996) and laboratory (Szczepanski \& Vala 1993, Hudgins \etal 1994, Hudgins \&
Allamandola 1995, Hudgins \& Sandford 1998) studies showed that ionized species have stronger C-C bands. 
On the other hand, dehydrogenated PAHs have weaker C-H bands (Pauzat \etal 1995, 1997). 
As we show in Sect. 4.4.1, the weaker C-H band emission in M17-SW can be explained if the hydrogenation 
fraction is significantly reduced.

Moreover, the AIB profiles show little spectral substructure even though most AIBs are fully resolved 
in our SWS data (the resolving power is $\lambda / \Delta\lambda = $ 200 to 500).
A direct comparison of the spectra displayed in Fig. 1 shows that the position and width of the major,
simple AIBs (at 3.3, 6.2, 8.6 and 11.3 \mum) are roughly the same. The decomposition of the spectra we discuss 
below primarily aims at quantifying this comparison and also at disentangling in a systematic way the profile of 
a given AIB from the other bands as well as from the underlying continuum due to hot small grains. \\       

Boulanger \etal (1998$b$) showed that the AIBs in CAM-CVF data can be decomposed
into Lorentz profiles and a linear underlying continuum. It must be emphasized,
however, that the AIBs are barely resolved in the CAM-CVF data (in particular
the 6.2, 8.6 and 11.3 \mum~ bands, see Table 2 in Boulanger \etal 1998$b$ and Fig. 3 of Cesarsky D. \etal
2000). The
SWS data at high spectral resolution presented here do not suffer from this limitation. 
The use of a lorentzian band shape implicitly assumes that the AIB profiles arise from the intrinsic
width of molecular transitions and/or resonances in small solid particles (\eg, Bohren \& Huffman 1983).

In the current paradigm, the AIBs result of the superposition 
of many vibrational bands produced by a population of interstellar PAHs with a wide range of sizes 
(molecules containing a few tens to a few hundreds carbon atoms, D\'esert \etal 1990, Schutte \etal 1993). 
In this respect, laboratory and theoretical studies on small PAH species teach us that the band shapes 
(position and width) of vibrational transitions {\em (i)} depend on the temperature of the molecule (Joblin \etal 
1995) and, {\em (ii)} vary from one species to another (in particular, the position of the vibrational bands 
depends on the size and symmetry of the molecule: Szczepanski \& Vala 1993, Hudgins \& Allamandola 1995 and 1998, 
Joblin \etal 1995, Langhoff 1996). The observed AIBs may actually result from a combination of 
these two effects. Elaborating on these studies, variability in 
the AIBs (changing band ratios, presence of substructure in the band profiles) from different interstellar 
sightlines is predicted as a consequence of a changing PAH population and/or different exciting radiation fields.
Specifically, the position and width of individual bands are expected to vary by a few to several tens of \cm_1
from one species to another and/or as a result of different emission temperatures.
Some changes in the AIB profiles have been observed towards \HII~ regions and reflection nebulae 
(Roelfsema \etal 1996, Verstraete \etal 1996, Cesarsky \etal 2000$a$, Uchida \etal 2000, Peeters \etal 2000 
in preparation). The case of the general (bright) interstellar medium is covered below with a quantitative comparison 
of the AIB profiles.

In this work, rather than establish the ``final'' AIB profile parameters, we aim at comparing on the same 
footing the individual AIB profiles under different excitation conditions. 
We have therefore decomposed the present spectra into Lorentz profiles and a 
modified blackbody as underlying continuum. We restricted ourselves to the minimum number of Lorentz
profiles required in order to produce a reasonable overall fit and a good representation of every 
individual AIB.

We used a classical gradient-expansion algorithm 
with analytical partial derivatives and performed the fit in the wavenumber space ($x=1/\lambda$ in \cm_1) 
over the 2.4-25 \mum~ wavelength range. In addition to the Lorentz profiles, 
the underlying continuum is fitted simultaneously. For the latter, we took a modified blackbody with an 
emissivity law proportional to $x$, the temperature and peak brightness of which were the free parameters.
The same set of fit parameters was used for all objects. Such a set was first fixed on the M17-SW 
spectrum which has a high signal-to-noise ratio and a good feature-to-continuum contrast. 
To fit the 2.4-25 \mum~ spectrum, twenty Lorentz profiles were necessary. 
Then, the parameter values of the Lorentz profiles and of the continuum in M17-SW were used as input to 
fit the other AIB spectra: very good fits were obtained by relaxing first the Lorentz amplitudes and 
blackbody continuum parameters, suggesting a complete and robust decomposition of the AIB spectrum.
After adjusting the amplitudes, the profile (position and width) and continuum parameters 
were fine-tuned, simultaneously, over restricted spectral ranges.
The centroids, widths and amplitudes of the lorentzian fits to the main AIBs are given in Table 2.
Our fit to the 5-25 \mum-AIB spectrum is shown in Fig. 2. The blackbody component of this decomposition
is consistent with the emission of warm dust in the mid-infrared and in particular its 
exponential decay (the Wien tail of the blackbody). The observed strong 
variability of the 20 \mum-continuum flux (very weak in NGC 2023 whereas strong in M17-SW and the Orion Bar)
then reflects the varying temperature of the warm dust component. 
As can be seen in Fig. 2, the underlying continuum contributes little below the AIBs. On the other hand, 
additional, colder blackbody type continua are required to fit the full SWS spectra out to 45 \mum. \\

We note that broad bands are required at about 1000 and 1450 \cm_1 in order to explain the continuum 
between the AIBs. These bands may not be associated with the AIBs but, for simplicity,
we assumed their profiles to be lorentzian. Their parameters are not well
constrained in our decomposition: the sole requirement is that 
the corresponding profiles are broad enough to reproduce the smooth continuum observed 
in these spectral regions. 
The fitted profiles of the neighbouring AIBs are somewhat sensitive to the widths 
adopted for the 1000 and 1450 \cm_1 bands: for instance, if the full width at half maximum (FWHM) of the 
1450 \cm_1 band is increased from 200 to 300 \cm_1, the 6.2 \mum-band has its FWHM reduced by 2.5 \cm_1 and 
its position redshifted by 0.8 \cm_1. In order to coherently compare the AIB profiles, we have fixed the 
width of these broad bands: namely, FWHM = 300 \cm_1 for the 1000 \cm_1-band and FWHM = 200 \cm_1 for the 
1450 \cm_1-band. At this stage, we can point out that combinations of PAH vibrational modes 
have been predicted to accumulate between 1000 and 2000 \cm_1 in a broad structure (Bernard \etal 1989).

\begin{table}[!t]
\begin{flushleft}
\caption[]{ Fit parameters of the AIB Lorentz profiles }
\leavevmode
\begin{tabular}[h]{lrrr}
\hline
\multicolumn{4}{c}{} \\
\multicolumn{1}{l}{AIB} &
\multicolumn{1}{c}{NGC 2023} &
\multicolumn{1}{c}{M17-SW} & \multicolumn{1}{c}{Orion Bar} \\
\multicolumn{4}{c}{} \\
\hline
\hline
\multicolumn{4}{c}{} \\
11.3 \mum  &  888.6 $^1$ &  889.9 &  889.7 \\
           &   20.8 $^2$ &   17.8 &   14.2 \\
	   &  877   $^3$ & 2292   & 7241   \\
Core       &  889.3 &  890.2 &  889.9 \\
           &   13.1 &   10.1 &   10.9 \\
	   &  564   & 1644   & 5928   \\
Red wing   &  881.9 &  880.3 &  880.5 \\
           &   29.6 &   30.4 &   30.0 \\
           &  384   &  915   & 1820   \\  
\\
\hline
\\
8.6 \mum   & 1164.0 & 1163.8 & 1161.5 \\
	   &   49.3 &   44.6 &   48.0 \\
	   &  355   &  780   & 1957   \\
\\
\hline
\\
7.8 \mum   & 1274.2 & 1273.5 & 1275.1 \\
	   &   70.2 &   67.6 &   54.2 \\
	   &  392   & 1494   & 2417   \\
7.6 \mum   & 1309.9 & 1312.6 & 1311.3 \\
	   &   22.9 &   28.6 &   25.6 \\
	   &  130   &  912   & 1760   \\
7.5 \mum   & 1328.1 & 1327.8 & 1329.7 \\
	   &   68.2 &   74.5 &   55.9 \\
	   &  466   & 1074   & 2022   \\
\\
\hline
\\
6.2 \mum   & 1608.1 & 1607.7 & 1609.7 \\
	   &   48.6 &   42.6 &   38.9 \\
	   &  550   & 1748   & 3428   \\
Core	   & 1608.5 & 1608.6 & 1610.5 \\
	   &   17.1 &   30.0 &   25.5 \\
	   &  224   & 1376   & 2369   \\
Red wing   & 1600.5 & 1586.2 & 1594.9 \\
           &   80.0 &   64.4 &   64.8 \\
	   &  338   &  546   & 1291   \\
\\
\hline
\\
3.3 \mum   & 3041.8 & 3039.1 & 3040.0 \\
	   &   43.0 &   38.8 &   40.4 \\
	   &   98   &  171   &  834   \\
\multicolumn{4}{c}{} \\
\hline
\end{tabular}\\
\end{flushleft}
$^1$ center in \cm_1 ($\pm$ 0.8 \cm_1, see text)\\
$^2$ width in \cm_1 ($\pm$ 2 \cm_1)\\
$^3$ amplitude in \megjysr
\end{table}

Also noteworthy is that two lorentzians are required to correctly reproduce the red 
wing asymmetry of the 1609 \cm_1 (6.2 \mum) and 890 \cm_1 (11.3 \mum) bands: these components are labelled
``core'' and ``red wing'' in Table 2. The feature centered around 1300 \cm_1 (the classical ``7.7 \mum-band'') 
shows 3 sub-peaks at about 1273, 1310 and 1328 \cm_1. In the following, we will call the narrow
1310 \cm_1 peak the ``7.6 \mum-band'' while the broader 1273 and 1328 \cm_1 components will be dubbed 
the ``7.8'' and ``7.5 \mum-bands'' respectively.
The observed 1310 \cm_1-feature 
has a narrow core and a broad blue wing which demands another component at 1328 \cm_1. 
A lorentzian is also required around 1204 \cm_1 (FWHM $\sim$ 70 \cm_1) to fill the gap between the 7.8 and 
the 8.6 \mum-features. In the case of Orion this band shifts to 1209 \cm_1 in order to reproduce the 
pronounced and extended blue wing of the 8.6 \mum-band.

In Fig. 3 we show the profile and fit of the 3.3 \mum~ band. The band shape is also
lorentzian and the continuum has the same functional form as in the 5-25 \mum~ region.
In other words, we fitted the 2.4-25 \mum~ continuum with a single blackbody component. 

The decomposition presented above is not unique: we also tried to use a multicomponent power-law 
continuum ($A/x^3$ + $B/x^2$ + $C/x$ + $D$ with $x=1/\lambda$ and where 
$A,B,C$ and $D$ are constant parameters) below the AIBs from 2.4 to 16
 \mum. Such a continuum gives a notable contribution below the AIBs, in particular, it replaces completely 
the 1000 \cm_1 broad band. However, a multicomponent power-law continuum has no straightforward physical 
interpretation and it cannot describe the 
strong rise of the spectrum beyond 16 \mum~ observed in M17-SW and the Orion Bar. Nor can it accommodate the 
weak 20 \mum-flux in NGC 2023. In any case, 
comparing the fits obtained with the two types of continua (power-law and blackbody), we found the positions 
and widths of the main AIBs characterized in Table 2 to vary by less than 1.6 and 4 \cm_1 respectively in a 
given object. 
Finally, using gaussian band shapes for the AIBs, we got as good a
decomposition (see also Boulanger \etal 1998$b$): yet the continuum under the
AIBs was stronger and more structured because of the weaker profile wings. We believe that the fit 
quality does not preclude any band profile: in fact, the SWS AIB spectrum is so rich that there are always 
enough parameters in the fit to accomodate any choice of band profile. 

\begin{figure}[]
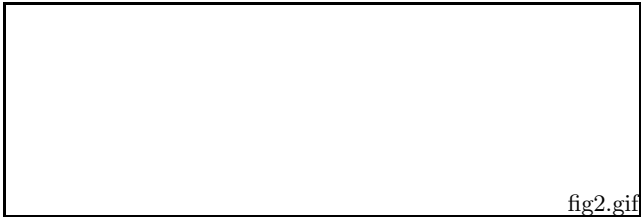


\framebox(240,80)[br]{fig2.gif}

 \caption{ The decomposition of the AIB spectra between 5 and 25 \mum. The resolving power 
           ($\lambda/\Delta\lambda$) of the data is 200 for NGC 2023 and 500 for M17-SW and the Orion Bar.
           Each Lorentz component is represented by a dotted line, the underlying continuum by a dashed line. 
	   The full fit is the solid line.
	  }
\end{figure}

We have thus determined in a coherent way the position and width of the strong, 
well-delineated AIBs at 3.3, 6.2, 7.6, 8.6 and 11.3 \mum~ with an accuracy of 0.8 and 2 \cm_1
respectively. 
Inspection of Table 2 shows that there are significant variations in the width of the 6.2 and 
11.3 \mum-bands (9.7 and 6.6 \cm_1 respectively) as well as in the position of the 3.3 and 
7.6 \mum-bands (2.7 \cm_1 in both cases). However, all these variations come within the accuracy 
range given above when the values of NGC 2023 are excluded. In this latter object, the poorer 
signal-to-noise ratio and resolving power (see Fig. 1 and Figs. 8 to 11) has degraded the band 
profiles: this is probably why some AIB parameters are singular in NGC 2023. On the other hand, 
we find that the position of the 8.6 \mum-band varies by 2.5 \cm_1 (Table 2) and that most of this 
variation is due to a redshift in the Orion Bar spectrum. As noted above, the 8.6 \mum-band in this object
has an extended blue wing: this spectral change is related to the redshift of the band itself. 
This result may point to the more profound modifications seen in this spectral range by Roelfsema \etal 
and Verstraete \etal (1996) towards more excited regions.

\begin{figure}[!t]
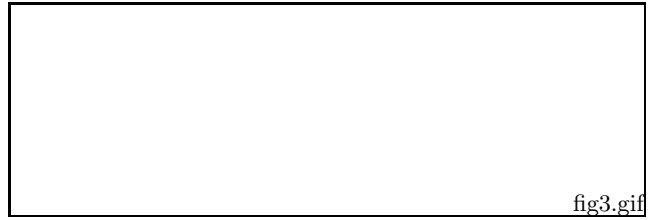

\framebox(240,80)[br]{fig3.gif}
 \caption{ Same as Fig. 2 in the region of the 3.3 \mum~ AIB.
 	   Note that the continuum has the same functional form than  
	   that already seen in Fig. 2. In the Orion Bar spectrum, the Pf$\delta$ hydrogen recombination 
	   line clearly visible on the top of the 3.3 \mum-band was masked to perform the fit. The Pf $\gamma$ 
           at 3.74 \mum~ and Pf $\epsilon$ at 3.04 \mum~ lines can also be seen in emission in the Orion Bar 
           spectrum.
	  }
\end{figure}

From the above discussion, we conclude that most AIB profiles (except the 8.6 \mum-AIB in the Orion Bar) 
do not vary to the accuracy of our spectral decomposition. 
At 3.3 \mum, a similar result was already obtained by Tokunaga \etal (1991) who compared the band profile of planetary 
nebulae and \HII~ regions (their Type 1 profile) 
with excitation conditions ($T_{eff}$ and $\chi$) similar to that of our sample (see Table 1). 
Roche \etal (1996) confirmed this result on a larger, lower-excitation sample of planetary nebulae. 
Similarly, Witteborn \etal (1989) showed that the 11.3 \mum-band profile is rather stable. 
Furthermore, the observed smoothness and invariance of the AIBs across a wide range of excitation conditions appears 
difficult to reconcile with the variability expected from laboratory results.\\

Having performed this mathematical parameterization of the SWS spectra, we are now able to extract the individual 
AIB profiles and to compare them in different objects and to model predictions. 
Yet, what is the physical meaning of this spectral decomposition ? We have chosen lorentzian band shapes and we 
represented most AIBs (except the 7.7 \mum) with one or two Lorentz profiles: this implicitly assumes that the 
AIBs arise from a few vibrational bands common to many PAHs and that most of the bandwidth arises from a single 
carrier. The invariance and smoothness of the AIBs is then naturally explained (Boulanger \etal 1998$b$). 
To verify this interpretation of the AIBs, the band parameters of Table 2 can 
be compared directly to laboratory or theoretical studies on relevant PAHs (large molecules 
containing more than 50 C-atoms as argued by Boulanger \etal 1998$b$). But there are other ways to look at the 
interstellar AIB spectrum which do not result in lorentzian bandshapes. 
For instance, simulations of astronomical spectra based on recent laboratory studies 
of small PAHs (Allamandola \etal 1999) have shown that the AIB spectrum may be decomposed by assigning a 
multiplicity of (species dependent) vibrational bands to each AIB. Another possibility to interpret the AIB spectrum 
is based on the laboratory work of Joblin \etal (1995) which shows that the vibrational bands of PAHs 
are broadened and redshifted as the temperature of the molecule is raised. In this picture, the profile 
of a vibrational band from a single molecule with a given internal energy has a Lorentz shape (the intrinsic profile) 
which only depends on the 
temperature of the molecule (and not on the species). The observed AIBs are then interpreted as the superposition of 
many Lorentz profiles corresponding to all the possible temperatures reached by a population of PAHs. \\

In the next section, devoted to modelling, we adopt and detail this latter view of the AIB spectrum. 
In this context, using the relationship between bandwidth and temperature established by Joblin \etal (1995), 
we note that the width of the observed 3.3 \mum-band (40 \cm_1) is well explained if the emitting
PAH has a temperature of about 1000 K. Such emission temperatures are in good 
agreement with what is expected for interstellar PAHs emitting during temperature fluctuations 
(see Sect. 4).  
Using the present spectral decomposition of the SWS data, we can now extract the individual AIB profiles 
and compare them to the predictions of a PAH emission model that uses the best available laboratory 
data on PAHs.

\section{Probing the PAH model}

To first order, the AIB profiles essentially remain identical 
while the exciting radiation hardens considerably: $T_{eff}$ = 23,000 K in NGC 2023 to 
45,000 K in M17-SW corresponding to the mean energy of the photons absorbed by a PAH 
of 6 and 9 eV respectively. Such contrasted internal energies
imply differences in the emission temperature of a few 100 K for a given molecular size. 
With this temperature change, band position shifts and band broadening of at least 10 \cm_1 are expected 
(Joblin \etal 1995). How can then the interstellar AIB profiles be so stable ? One possibility 
is that some process keeps the temperature distribution of interstellar PAHs essentially 
the same. \\

To investigate this issue, we have computed the temperature distribution and
associated infrared (IR) emission of an interstellar PAH population using a modified version of the model 
described in Pech \etal (2000 hereafter PJB) where we study in detail the temperature fluctuations and the 
temperature distribution of PAHs. Improving on former work (L\'eger \etal 1989$b$, Schutte \etal 1993, 
Cook \& Saykally 1998), this model takes 
into account the full spectral distribution of the exciting radiation and implements the recent results of Joblin 
\etal (1995) on the {\em temperature dependence} of the band profiles of {\em neutral} PAHs. 
In addition, it must be emphasized that the PAH IR emission {\em cross-section} of the PJB model was derived from 
studies on PAH {\em cations}: indeed, we show in Sect. 4.4.1 that singly ionized PAHs 
reproduce better the astronomical AIB spectra confirming previous work (Langhoff 1996, Cook \& Saykally 1998, 
Allamandola \etal 1999, Hudgins \& Allamandola 1999). This observational requirement is supported by theoretical 
studies on the charge state of interstellar PAHs for physical conditions comparable to those of our sample 
of objects (Bakes \& Tielens 1994, Dartois \& d'Hendecourt 1997).
We first briefly describe the equations giving 
the PAH cooling curve and emission, compute the temperature distribution of interstellar PAHs and then
compare our model PAH emission spectrum to the data. For the sake of clarity, we limit the model-to-data
comparison to NGC 2023 and M17-SW which are the extreme cases of our data sample and we focus on the well-defined 
3.3, 6.2, 8.6 and 11.3 \mum-bands. 
In order to emphasize the temperature fluctuations of PAHs 
and the resulting temperature distribution, we have written the model equations in terms of cross-sections rather 
than Einstein $A$-coefficients: this formulation is however strictly equivalent to that of PJB (see, \eg, 
Schutte \etal 1993, Eq. 15).

\subsection{Emission by PAHs}

In the interstellar medium, a PAH containing $N_C$ carbon atoms absorbs radiation from the surrounding 
stars at a rate:

\begin{equation}
R_{abs}({\rm photons\;s}^{-1})=\int^\infty_0 \sigma^{a}_{\nu}\;
\frac{F_{\nu}}{h\nu}\; d\nu
\end{equation}

\noindent with $\sigma^{a}_{\nu}$ the visible-ultraviolet (UV) absorption cross-section of PAHs 
and $F_{\nu}$ the incident stellar flux assumed to be a diluted blackbody with the dilution factor and
effective temperature given in Table 1. The absorption cross-section is proportional to $N_C$; it also  
has a cut-off (due to the electronic transitions) in the visible to near-IR range whose wavelength increases 
with $N_C$ (Verstraete \& L\'eger 1992). The mean photon energy absorbed by a molecule is given by:

\begin{equation}
E_{abs} = \frac{P_{abs}}{R_{abs}} \quad{\rm with} \quad
P_{abs}=\int^\infty_0 \sigma^{a}_{\nu}\;F_{\nu}\; d\nu.
\end{equation}

\noindent $P_{abs}$ is the power absorbed per molecule. In the standard interstellar radiation field of Mathis 
\etal (1983), we find that a PAH with $N_C$ = 25 or 500 carbon atoms absorbs $2.4\times 10^{-27}$ 
or 4$\times 10^{-27}$ Watt per C-atom with $E_{abs}$ = 4.8 or 2.3 eV, respectively. The values for $N_C$ = 25 are 
in good agreement with the laboratory results of Joblin \etal (1992) on small PAHs. 

As shown by L\'eger \etal (1989$a$) and Schutte \etal (1993), the IR emission of a 
PAH is well treated within the thermal approximation and using 
Kirchhoff's law. In the following, we will discuss the complete spectrum emitted by PAHs in terms of 
the spectral energy distribution (SED) noted $S(\nu)$ which is equal to 
$\nu I_{\nu}$. The SED emitted by a single 
molecule containing $N_C$ carbon atoms is thus: 

\begin{eqnarray}
&S(\nu',&N_C) = 4\pi\,R_{abs} \nonumber\\
        & \times & \int^\infty_0 {\cal P}_{\nu} \;
\int^{t_e}_0 \nu'\;\sigma^{e}_{\nu'}\;B_{\nu'}\left( T_c(t,\nu,N_C)
\right) \; dt \;d\nu 
\end{eqnarray}

\noindent where 

\begin{equation}
{\cal P}_{\nu} = \frac{\sigma^{a}_{\nu}} {R_{abs}} \times \frac{F_{\nu}}{h\nu}
\end{equation}

\noindent is the probability density for absorption
of a photon of energy $h\nu$, $\sigma^{e}_{\nu'}$ the emission cross-section in the
IR, $t_e$ is the time at the end of cooling (see Sect. 4.2) and $T_c(t,\nu,N_C)$ is 
the cooling curve of the molecule. 
In the following, the frequency of photons emitted in the IR 
will be labelled $\nu'$ and that of stellar photons absorbed in the
visible-UV range will be written $\nu$. 

The PAH IR emission cross-section is defined as in PJB. The band profiles have a lorentzian shape
and at 3.3, 6.2, 8.6 and 11.3 \mum, the band position and width follow temperature dependent laws. 
Each Lorentz profile is normalized to the peak value of the band cross-section, $\sigma_{\nu}$ and 
we assume that the integrated cross-section (or Einstein $A$-coefficient, see Schutte \etal 1993) of each band profile 
is conserved with temperature. As already noted above, the resulting IR cross-section describes the properties of singly-ionized PAHs.
To match the observed AIB, the 8.6 \mum-band had to be multiplied by 3. Because the available laboratory results do not account for the 
complex shape of the observed 7.7 \mum-band (see Sect. 3), we define an empirical 7.7 \mum-band with a shape derived from the 
observations (profile centre at 1300 \cm_1 and FWHM = 113 \cm_1 from the present data) and 
with the (laboratory determined) Einstein $A$-coefficient of PJB. This profile of the 7.7 \mum-band is furthermore assumed temperature 
independent. We also added the 16.4 \mum-AIB detected by ISO towards many sightlines (Moutou \etal 2000,
van Kerckhoven \etal 2000). This band is often 
associated with the ``classical'' 3 to 13 \mum-AIBs and its intensity is not related to the hot dust
continuum appearing in more excited objects (see Fig. 1), \ie, as for the other AIBs.
We represent the 16.4 \mum-band with a Lorentz profile centered at 609 \cm_1, with FWHM=6 \cm_1
(both being observational values, see Moutou \etal 2000) and an Einstein $A$-coefficient of $3.1\times 10^{-2}$ s$^{-1}$ per carbon atom 
which corresponds to the average value of the laboratory measurements given in Moutou \etal (1996, Table 3). 
As we will see below, the intensity of the 16.4 \mum-band can constrain the power-law index of the PAH 
size distribution. At longer wavelengths, we adopted the far-IR ($\lambda\geq$ 20 \mum) cross-section 
of Schutte \etal (1993).

The size distribution of PAHs is defined by $dN=B\,N_C^\beta\,dN_C$, {\it i.e.}, the number of PAHs with a 
number of C-atoms between $N_C$ and
$N_C\,+\,dN_C$. Writing the size distribution with respect to $N_C$ frees us of
any assumptions on the PAH geometrical shape
\footnote{Using the grain radius, $a$, as a variable the size distribution
reads $d{\cal N}\sim a^\alpha\,da$. We thus have the relationship $\beta=(\alpha-d+1)/d$,
where $d$ is the dimension of the grain: $d=2$ for planar grains and $d=3$ for spherical grains.
For instance, assuming PAHs are planar ($a=0.9\sqrt{N_C}$) a value of -3.5 for $\alpha$ 
corresponds to $\beta=-2.25$.}. The number of H-atoms per
molecule is assumed to be $N_H=f_H\times\sqrt{6N_C}$ (Omont 1986, in the case of symmetric PAHs) with $f_H$ the 
hydrogenation fraction of a molecule: if $f_H$=0 the molecule is completely dehydrogenated, if  $f_H$=1 the 
molecule is fully hydrogenated. Unless otherwise stated, we will assume $f_H$=1.
The mass distribution is then straightforwardly found as $dN\times (m_C\;+\;m_H\times f_H\sqrt{6/N_C})$ 
where $m_C$ is the carbon atomic mass and $m_H$ the atomic mass of hydrogen. The 
normalization factor, $B$, is determined from the PAH abundance in terms of the 
fraction of interstellar carbon ([C/H]$_{ISM}=2.6\times 10^{-4}$, Snow \& Witt 1996) locked up 
in PAHs. Finally, the emergent SED emitted by PAHs is 

\begin{equation}
S^e(\nu')=\int^{N_{max}}_{N_{min}} S(\nu',N_C)\; dN(N_C)
\end{equation}

\noindent with $N_{min}$ and $N_{max}$
the smallest and largest number of C-atoms per molecule respectively.
We also define the total luminosity emitted by PAHs as
$L_e = \int^{\infty}_0 S^e_{\nu'} \; d\,ln\,\nu'$.

\subsection{The infrared cooling of PAHs}

Upon absorption of a UV-photon, the PAH temperature is raised to a peak value,
$T_p$. The molecule then cools rapidly by the emission of IR vibrational mode photons.
The peak temperature $T_p$ is found from first principles:

\begin{equation}
U_f-U_i = \int^{T_p}_{T_i} C(T)\,dT
\end{equation}

\noindent where $C(T)$ is the heat capacity of PAHs as given by Dwek \etal (1998, Eq. A4 of their 
Appendix), $U_{i,f}$ are the initial and final molecular internal energies and
${T_i}$ the PAH temperature prior to photon absorption or the mean molecular 
temperature between two photon absorptions. 
We assumed that
$U_f>>U_i$ with $U_f=h\nu$ the energy of the absorbed photon. Furthermore, we took 
$T_i$ to be the PAH temperature at the end of cooling (see the definition below). The heat capacity, 
as a function of temperature, is approximated by a polynomial in $T$; it is also proportional to $N_C$
but independent of $N_H$, \ie, the contribution of the hydrogen atoms to $C(T)$ is assumed to be 
always the same whatever the size of the molecule (Dwek \etal 1998). 
In this case and for a given energy per C-atom in the molecule ($h\nu/N_C$), $T_p$ is fixed.\\

The cooling curve $T_c(t,\nu,N_C)$ is obtained by numerical integration ({\em vs.} temperature) 
of the energy conservation relationship during cooling, 

\begin{equation}
P_e\left(T_c\right)\,dt = - C(T_c)\,dT
\end{equation}

\noindent where $P_e$ is the total power emitted at all wavelengths by the molecule and 
$dT$ the temperature drop during the time $dt$.
This simple differential equation determines a family of cooling curves, $T_c$, as a solution,
each cooling curve being completely defined by the knowledge of $N_C$ and of the initial condition, 
\ie, $T=T_p$ at $t=0$. 

The power emitted by a PAH at temperature $T$ is: 

\begin{equation}
P_e(T)=\int^\infty_0 \sigma^{e}_{\nu'} \times 4\pi B_{\nu'}(T)\, d\nu'.
\end{equation} 

Cooling curves were computed on a grid of 50 points with constant temperature intervals. 
Under these conditions, we found that the energy conservation requirement:

\begin{equation}
\int^{\infty}_0 P_e(T_c(t)) \; dt = h\nu
\end{equation}

\noindent is always met within a few percent.

Each molecule sees its cooling interrupted by the absorption of the next photon which 
is a stochastic process (D\'esert \etal 1986). Moreover, 
gas-grain exchanges must be taken into account (Rouan \etal 1992, Draine \& Lazarian 1998) 
to reliably estimate the temperature of a PAH in the low temperature part of the cooling curve. 
In fact, the emission from 
the cool (but long-lasting) tail of the cooling curves will peak in the far-IR/submillimeter, far away 
from the 3 to 20 \mum-range on which we focus here. The 
detailed treatment of this problem is outside the scope of this paper and we adopt the 
following simple criterion to truncate our cooling curves.
The mean time elapsed between two photon absorptions is $1/R_{abs}$: it is for instance 6.9 and 2.6 hours 
in the radiation field of NGC 2023 and M17-SW respectively and for a molecule with $N_C$ = 20.
Consequently, all cooling curves have been truncated at $t_e=1/R_{abs}$ to
account for the repeated absorptions of UV photons by a molecule; $t_e$ decreases as $N_C^{-1}$ 
because $R_{abs}\sim N_C$. We thus define the lowest 
temperature of a PAH heated by a photon of energy $h\nu$ as $T_i=T_c(t_e,\nu,N_C)$, \ie, equal 
to the temperature at the end of the cooling.
For instance, in the radiation field of NGC 2023 small PAHs 
($N_C=20$) eventually reach temperature as low as 15 K at time $t_e$ whereas bigger PAHs 
($N_C=500$) remain at around 35 K.

\begin{figure}[!ht]
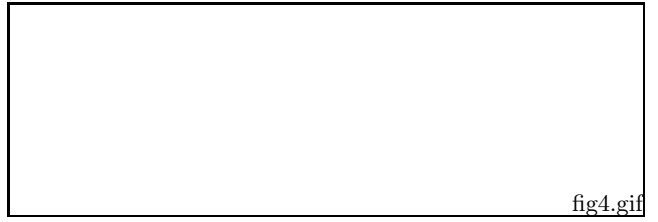

\framebox(240,80)[br]{fig4.gif}
 \caption{ Cooling curves for various energy per carbon atom (the zero time point has 
been excluded). 
The numbers in parenthesis give the value of $E_C=h\nu/N_C$ in meV/C.
The corresponding $(h\nu,N_C)$ combinations were as follows: 
$(a)$ $E_C$=680 meV/C, solid line is ($h\nu$=13.6 eV, $N_C$=20), 
$(b)$ 453 meV/C, solid = (13.6, 30) and dot-dash = (9.06, 20), $(c)$ 272 meV/C,
solid = (13.6, 50) and dot-dash = (5.44, 20), $(d)$ 136 meV/C, solid = 
(13.6, 100) and dot-dash = (2.72, 20), (e) 27.2 meV/C, solid = 
(13.6, 500) and dot-dash = (0.54, 20)
}
\end{figure}

In Fig. 4, we show
the cooling curves obtained for various energies per C-atom $E_C=h\nu/N_C$. As expected, the
peak temperatures are identical for a given $E_C$. We note that 
the $T_c$-curves do not change much for different $N_C$-values while $E_C$
remains fixed: in fact if the emission cross-section were only proportional to $N_C$ 
all the cooling curves with ($h\nu$,$N_C$) combinations yielding the same 
$E_C$ would merge. Cooling curves with larger $N_C$ have higher temperatures
because emission in the C-H modes is less important (in a fully 
hydrogenated PAH we have $N_H/N_C=\sqrt{6/N_C}$). The small variation in the $T_c$-curves at fixed $E_C$ 
then reflects the decrease of the emission cross-section per carbon atom with increasing size.

\subsection{The temperature distribution of interstellar PAHs}

As noted above, the observed similarity of the AIB profiles in our sample means that
the range of PAH emission temperatures does not vary much: to test this idea, we determine 
here the temperature distribution of PAHs. Moreover, as we show below, the effect of the 
size distribution and exciting radiation field parameters are illustrated in a synthetic fashion 
in the temperature distribution. \\

The observed AIB spectrum results principally from the superposition of many blackbodies
at different temperatures (Eqs. 3 and 5). Following Eq. 3 it is clear that
a given blackbody will have a significant contribution to the emergent 
spectrum if its temperature $T$ is high and/or if it remains a long time 
at that temperature: a PAH with a temperature between $T$ and 
$T+dT$ (corresponding to the time interval $[t,t+dt]$) will thus contribute to the 
total emitted luminosity $L_e$ (see the end of Sect. 4.1) with a weight proportional to $P_e(T)\times dt$ 
or the fraction of the total available energy, $h\nu$ (the energy of the absorbed photon), that is dissipated 
between $t$ and $t+dt$. 
This definition of the temperature weight is different from previous work 
(Draine \& Anderson 1985, D\'esert \etal 1986) which only considered the time that
a given grain spends in the temperature interval $[T,T+dT]$, namely $dt/dT$: 
the present definition which takes into account the emitted power actually reflects
the contribution of a given temperature to the final spectrum. \\

From Eq. 7, we see that the temperature weight in the total luminosity is simply $C(T)\times 
dT$. Taking into account the probability distribution of the exciting photons and the fraction 
of PAHs containing $N_C$ carbon atoms, we define the temperature weight $w_T$ as:

\begin{equation}
w_T =  \frac{C(T_c)\;\delta T} {h\nu} \times {\cal P}_\nu\;\delta \nu \times 
\frac{N_C^{\beta+1}\;\delta\,ln\,N_C} {N_{PAH}}
\end{equation}

\begin{figure}[!ht]
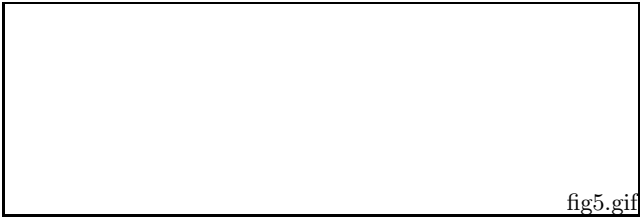

\framebox(240,80)[br]{fig5.gif}
 \caption{ The distribution of PAH temperatures $T\times p(T)$ {\bf normalized to 1 at peak value}. In the panels $a)$, $b)$ 
           and $c)$, we used the parameters for the radiation field of NGC 2023 (see Table 1). 
	   The fixed parameters of each plot are given in brackets.
           In $a)$, we vary $N_{max}$ and the triple-dot dash line shows the effect of assuming that
           all molecules absorb the same photon ($E_{abs}$ = 6.2 eV). Variations of $N_{min}$ are explored 
	   in $b)$. The power-law index, $\beta$, is changed in $c)$. In $d)$, we explore the effect of 
           $T_{eff}$ while keeping the dilution factor of NGC 2023 (see text) and $N_{min}$ = 20. The solid line 
           corresponds to the case of the NGC 2023 radiation field. The dot-dash line in $d)$ shows the case of the M17-SW 
           radiation field 
           with $N_{min}$ = 30, the value required to fit the observed AIB profiles (see Sects. 4.4.1 and 4.4.2).
	  }
\end{figure}

\noindent where $N_{PAH}$ the total number of PAHs and 
$\delta T$ is the temperature bin in $T_c$, $\delta\nu$ the frequency bin of the 
exciting radiation field and $\delta\,ln\,N_C$ the logarithmic bin of the size distribution. All these 
bins have been taken as fixed. With this definition we have $\sum_{(T,N_C,\nu)} w_{T} = 1$. \\

To obtain the PAH temperature distribution in a given exciting radiation field, we build the 
histogram of all cooling curves for all molecular sizes in temperature bins of constant size 
($\Delta T$ = 100 K), each temperature $T$ being given the weight $w_T$.
The density value in the histogram is then simply $\sum^{T+\Delta T}_{T'=T} w_{T'}$.
The distribution per temperature interval, $p(T)$ is then found by normalizing the histogram so that 
$\int_0^{\infty} p(T)dT$ = 1.

\begin{figure}[!ht]
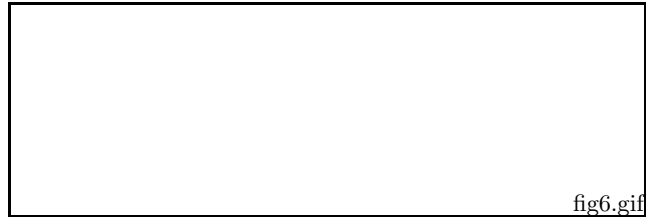

\framebox(240,80)[br]{fig6.gif}
 \caption{ The spectral energy densities corresponding to the temperature distributions shown in Fig. 5. 
           All the curves have been normalized to 1 at the peak value of the 7.7 \mum-band (see text). 
           In panels $a)$ to $c)$, the exciting radiation field is that of NGC 2023 and we vary the parameters 
           of the PAH size distribution with respect to some reference values, namely, $N_{min}$ =20, 
           $N_{max}$ = 200 and $\beta$ = -2.25. In $d)$, we look at the effect of changing the effective temperature
           of the exciting radiation field while the dilution factor is always equal to that of NGC 2023. 
           In each panel, the parameters kept constant are given in brackets.           
	  }
\end{figure}

From our weight definition, $p(T)dT$ is the fractional contribution of all PAHs with temperatures
in the range $[T,T+dT]$ to the total emitted luminosity $L_e$. \\

The PAH temperature distributions, normalized to 1 at their peak values to facilitate the comparison, are shown in Fig. 5: 
they all present a maximum between 300 and 800 K. 
The shape of these distributions can be understood as follows. Assuming that all molecules absorb a photon of energy 
$h\nu$, the temperature weights are proportional to $C(T)\times N_{C}^{\beta+1}$. At low $T$, all the $C(T)$-curves corresponding 
to the different $N_C$'s sum up to give $w_T$. As the temperature increases, the largest molecules do not contribute anymore to $w_T$ 
because they are too cold: this fade out more than compensates for the rise of $C(T)$ with the temperature and produces a maximum in the 
temperature distribution. 

Fig. 5 illustrates the sensitivity of the PAH temperature distribution to the size distribution parameters,
$N_{min}$, $N_{max}$ and $\beta$ (Fig. 5$a$ to 5$c$) and to the hardness of the exciting radiation 
field, parameterized as $T_{eff}$ (Fig. 5$d$). The associated changes in the AIB spectrum are shown in Fig. 6.
In all these figures (except the M17-SW case in Fig. 5$d$), we used the same dilution factor for the radiation field 
($W_{dil}=2.66\times 10^{-13}$, corresponding to NGC 2023): indeed, as long as one is in the regime of temperature fluctuations, 
the spectral distribution of the PAH emission spectrum does not depend on the radiation field intensity (Sellgren 1984) while its 
absolute level scales with it (see Eq. 3). Moreover, for the purpose of comparison, all the spectra of Fig. 6 have been normalized 
to 1 at the peak value of the 7.7 \mum-band because its profile is independent of the PAH temperature (see Sect. 4.1).

In Fig. 5$a$, we consider the effect of taking the full spectral distribution of exciting photons 
instead of the same mean absorbed photon energy, $E_{abs}$ (\ie, ${\cal P}_{\nu}=\delta[h\nu-E_{abs}]$).
In the following, we will refer to the {\em $E_{abs}$-approximation} whenever it is assumed that each molecule absorbs the 
same photon of energy $h\nu=E_{abs}$ at the rate $R_{abs}$ in the given radiation field. The $E_{abs}$-approximation
only affects the high temperature tail of the distribution. As we will see in Sect. 4.4.2, this tail matters for 
the width of the 3.3 \mum-band. Nevertheless, the rest of the AIB spectrum is little affected by the $E_{abs}$-approximation. 
$N_{max}$ has a weak influence on the temperature distribution (Fig. 5$b$) and its impact, which is most noticeable at long 
wavelengths ($\lambda \geq$ 10 \mum, see Fig. 6$a$), is easily drowned by small changes of $\beta$, the index of the 
power-law size distribution of PAHs (see Figs. 5$c$ and 6$c$). When $\beta$ decreases, the contribution of the smallest PAHs 
becomes more important and the cold tail of the temperature distribution is somewhat reduced (Fig. 5$c$): this leads to an enhanced 
3.3/11.3 \mum-band ratio and to somewhat broader bandwidths (Fig. 6$c$ and Figs. 8 to 11). 
Note that, due to this broadening, the band peak value diminishes 
(see the 6.2 \mum-band in Fig. 6$c$) because the band integrated cross-section is conserved with temperature. 
Both $N_{min}$ and $T_{eff}$ (which determines $E_{abs}$, here $T_{eff}$ = 10,000 to 45,000 K corresponds to $E_{abs}$ = 4 to 9 eV) 
affect the hot tail of $p(T)$ (Figs. 5$b$ and 5$d$) which merely reflects the hottest part of the cooling curve
(Fig. 4). 
This behaviour of $p(T)$ follows from the fact that the PAH peak temperature only depends on the energy content per carbon atom in a molecule, 
$E_C$ (see Sect. 4.2) and that a typical value for $E_C$ is $E_{abs}/N_{min}$ ($N_{min}$ is close to the mean value of $N_C$ because of the 
steep power-law size distribution). In fact, the temperature distribution and the shape of the emission spectrum are similar for different 
$(E_{abs},N_{min})$-couples yielding the same energy per carbon atom. As expected, an increase of $N_{min}$ leads to colder PAHs (Fig. 5$b$) 
which contribute more to the long-wavelength bands (Fig. 6$b$) while an increase of $T_{eff}$ has the reverse effect (see Figs. 5$d$ and 6$d$). 
We pointed out earlier that the PAH temperature distribution must remain unchanged in order to explain the similarity of the observed AIB profiles:
we show in Fig. 5$d$ that this condition is fulfilled if $N_{min}$ is raised along with $T_{eff}$. Specifically, an increase of $T_{eff}$ from  
23,000 K (the case of NGC 2023) to 45,000 K (the case of M17-SW) requires $N_{min}$ = 20 to 30, respectively. 
In the next sections, we show that this requirement allows to consistently reproduce the overall AIB spectrum as well as the individual band 
profiles.

\subsection{Modelling the AIB spectrum}

We now present the PAH IR-emission as computed from the formalism described
above in the case of NGC 2023 and M17-SW. We selected these two objects because
they present the largest contrast in the effective temperature of the exciting 
radiation field: we therefore expect these spectra to provide the strongest constraints
on our model.

\subsubsection{The complete spectra}

\begin{figure}[!ht]
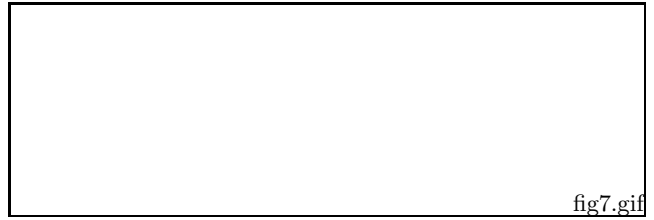

\framebox(240,80)[br]{fig7.gif}
 \caption{ The spectral energy density in Watt per hydrogen for our sample of objects from the SWS data (dotted curve)
	   and assuming a column density of 1.8 10$^{21}$ cm$^{-2}$.  The resolving power 
           ($\lambda/\Delta\lambda$) is 200 for NGC 2023 and 500 for M17-SW.
           The model PAH emission is the dot-dash line. The solid line shows the model complete spectrum 
           which includes the broadband continuum described in Sect. 3.
	   We took $N_{max}=200$ and $\beta=-3.5$ for both objects. For NGC 2023, we take
           $N_{min}$ = 20, $f_H$ = 0.8 and 10\% of the interstellar carbon in PAHs while we 
           require $N_{min}$ = 30, $f_H$ = 0.5 and 8\% of the interstellar carbon in PAHs for M17-SW
           (see text). 
           In the upper panel, the dashed line shows the case of neutral PAHs from our model with 
           $N_{min}=20$, $N_{max}=200$, $\beta= -3.5$, $f_H=1$.  For clarity, the SED of neutral PAHs has been normalized 
           to the peak value of the 11.3 \mum-band in the SED of ionized PAHs (dot-dash line). This normalization amounts to a PAH 
           abundance representing 6\% of the interstellar carbon. 
	  }
\end{figure}

Having determined the temperature distribution of a population of PAHs, we can
now compute the corresponding IR emission using eqs. 3 and 5. 
The PAH size distribution parameters were constrained as follows. 
We note that the overall shape of the 3-13 \mum~ AIB spectrum which is dominated by emission from molecules at 
$T\geq$ 400 K is not much affected by changes in $N_{max}$ or $\beta$ (see Figs. 6$a$ and 6$c$).
For simplicity, we fix from now on $N_{max}=200$. 
First, $N_{min}$ is fixed so as to reproduce 
the observed 3.3/11.3 \mum-band ratio in a given exciting radiation field (described by $T_{eff}$): 
we require $N_{min}=20$ in NGC 2023 while $N_{min}=30$ in M17-SW. The hydrogenation fraction, $f_H$, is 
constrained from the 7.7/11.3 \mum-band ratio: a good match to the observed band ratios is obtained with $f_H$ = 0.8 
for NGC 2023 and $f_H$ = 0.5 for M17-SW.
The index of the power law size distribution, $\beta$, can be constrained from the requirement that the 
predicted 16.4 \mum-band (on top of the broadband continuum) matches the observed band: 
we find that $\beta$=$-3.5$ reproduces the 16.4 \mum-band well in both objects
\footnote{ This result is not sensitive 
to the amount of cooling occurring in the far-IR ($\lambda \leq 20$ \mum): multiplying or dividing our 
far-IR cross-section by 10 did not alter much our match of the 16.4 \mum-band. }.
Unfortunately, individual PAH species show a large spread in the integrated 
cross-section of the 16.4 \mum-band (see Table 3 of Moutou \etal 1996) which does not constrain $\beta$: 
specifically, to match the observed 16.4 \mum-band we require $\beta=-2.25$ for the lowest 
cross-section value while $\beta=-6$ for the largest cross-section value.
More laboratory work on PAHs is warranted in order to understand the trend of the 16.4 \mum-band 
profile with respect to $N_C$ and to the various isomer states.
As discussed by PJB, an additional constraint on $\beta$ is provided by the intensity and width of the AIB 
profiles (see the next section).  
The $\beta$-value can then be adjusted in order to 
match the peak position and the FWHM of the 3.3 \mum-band (see Sect. 4.4.2): in fact, the value of -3.5 found 
from the 16.4\mum-band gives a good fit in the case of NGC 2023 and M17-SW (see Fig. 8). 

At this point, it is interesting to note that that the 6-to-13 \mum~ part of the spectrum is produced by 
molecules with emission temperatures between 400 and 1000 K. During temperature fluctuations, 
most PAHs go through these temperatures:
this is why the maximum of the temperature distribution (Fig. 5) always falls in this range.
In fact, the emission of any PAH with $N_C\leq 160$ shows ratios of band 
peak values which are quite comparable to the observed 6-13 \mum~ AIB spectrum (see PJB and Draine \& Li 2001). On the other hand, 
the match of the observed band position and width in the present data sample and in CAM-CVF spectra of similar 
interstellar regions requires small molecules (see next section). However, taking into account the 
full size distribution of PAHs will add emission mostly at 3.3 \mum~ (from the smaller species, see Fig. 6$b$) and in the far-infrared 
(from the larger species, see Fig. 6$a$) without changing much the 6-13 \mum~ region. Thus, in the framework of the present 
model, the observed invariance of the AIB spectrum in CAM-CVF data (6-16 \mum) implies that the AIB 
carriers are small enough to undergo large temperature fluctuations (at least 400 K) during which 
they emit in this wavelength range.

Our best fit SED's are displayed on Fig. 7: we assumed a total column density of 
1.8$\times 10^{21}$ cm$^{-2}$ (\ie, 1 visible magnitude of extinction) for both NGC 2023 and M17-SW with 
10\% and 8\%, respectively of the interstellar carbon in PAH cations (using [C/H]$_{ISM}$ = 2.6$\times 10^{-4}$,  
Snow \& Witt 1996). The hydrogenation fraction of PAHs was taken to be 80\% for NGC 2023 and 50\% for M17-SW. 
To compare in a meaningful fashion the predicted PAH emission to the observed spectrum, the broadband continuum described 
in Sect. 3 has been added to our model spectrum. This continuum consists of the modified blackbody along with 
the 1000 and 1450 \cm_1 broad profiles (in M17-SW an additional broad profile at 600 \cm_1 is required, 
see Fig. 2). All the main AIBs (at 3.3, 6.2, 7.7, 8.6 and 11.3 \mum) are matched within 20\%. 
We note that the model fails to reproduce the 12.7 \mum-band in M17-SW: this may be due to an ill-definition of the 
underlying broadband continuum in this complex spectral region (see Fig. 2).
The AIB spectrum corresponding to the neutral 
PAHs (we took the IR cross-sections from L\'eger \etal 1989$b$) is shown in the case of NGC 2023 with a PAH 
abundance corresponding to 6\% of the interstellar carbon, with $N_{min}=20$, $N_{max}=200$ and $\beta=-3.5$.
The $N_{min}$-value was fixed so as to reproduce the position and width of the 3.3 \mum-band (see next section) 
and it is the same as for the cations. 
In the case of neutral PAHs, the C-C/C-H band ratio is reduced by about one order 
of magnitude with respect to the cations. Matching the observed AIB with neutral PAHs would require a very low 
hydrogen coverage ($f_H\leq 0.1$) which would produce a strongly discrepant 8.6 \mum-band and also would be at 
odds with theoretical predictions on the hydrogen coverage of PAHs (Allain \etal 1996). \\

In the following, 
we compare in detail the profiles of the observed 3.3, 6.2, 8.6 and
11.3\mum-AIBs to our model results: we restrict our comparison to these bands
because they are well defined in the data and involve a small number of Lorentz fit
components.

\subsubsection{Individual AIB profiles}

To do a detailed model-to-data comparison, we now study the individual AIB profiles. 
Due to anharmonic couplings, the line profiles of PAH vibrational transitions are 
redshifted and broadened as the temperature increases and both follow a linear law (Joblin \etal 1995). 
The temperature dependent broadening laws of PJB used in this modelling do not include the rotational width.
The centroid and width of each band profile are given by $\nu(T)=\nu_0+\chi_c\times T$
and $\Delta\nu(T)=\Delta\nu_0 + \chi_w\times T$ respectively.
The coefficients of the linear $T$-laws have been measured in the laboratory on a restricted 
sample of small ($N_C\leq 32$), neutral PAHs: we assume that the temperature dependence of the 
vibrational band profile are the same for all PAH sizes as well as for cations and neutrals. 
To match the position of the observed AIB profiles, we had to slightly change the band centers at 0 K: 
namely, for the 3.3, 6.2, 8.6 and 11.3 \mum-bands we took $\nu_0$= 3079 (3076), 1636 (1627), 1169 (1141) 
and 899 (896) \cm_1 where the laboratory values are in parenthesis. Such shifts are reasonable
in view of the laboratory results on the spread in band positions for different species.\\

\begin{figure}[!ht]
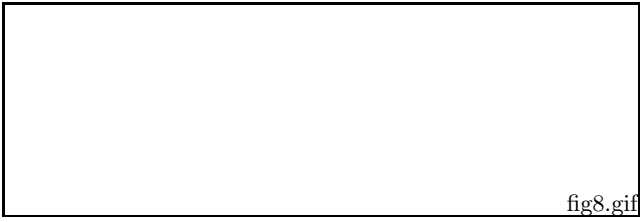

\framebox(240,80)[br]{fig8.gif}
 \caption{ The normalized profile of the 3.3 \mum-AIB as seen by ISO-SWS compared to our model emission 
           profile also normalized to one (see text). Note that the 3.4 \mum-band as well as the underlying plateau 
           (as represented in Fig. 3) have been subtracted from the 3.3 \mum-AIB.
	   The resolving power of the data ($\lambda/\Delta\lambda$) is 200 for 
           NGC 2023 and 500 for M17-SW. For all the model profiles, we took $N_{max}=200$.
           In $a)$, the SWS spectrum (from which the continuum and the
           contribution of all bands other than the 3.3\mum~ have been subtracted) is 
           the dotted curve, the result of our {\em model best fit} with $N_{min}=20$ and $\beta=-3.5$ 
           is the solid line and the dashed line represents the case where $\beta=-2.25$ (while $N_{min}=20$).            
           In $b)$, we show the contribution of the small ($20 \leq N_C \leq 35$, dashed line) and 
	   large ($35 \leq N_C \leq 200$, dot-dashed line) PAHs to the model best fit of NGC 2023. 
	   In $c)$, the observed 3.3 \mum-band profile of M17-SW (dotted curve) is shown as 
	   well as our {\em model best fit} (solid line) with $N_{min} = 30$ and $\beta=-3.5$. The model 
	   profile for $N_{min}=20$ (while $\beta=-3.5$) is shown as the dashed line. The dot-dashed line shows a 
           model using the $E_{abs}$-approximation ($E_{abs}=9.1$ eV, see Sect. 4.3) with $N_{min} = 30$ and $\beta=-3.5$.
	  }
\end{figure}

In order to highlight the match in band width and position, 
our model profiles are compared to the observed AIBs in a normalized fashion. For all the results of best fit models
represented, the peak absolute intensities do not deviate by more than 20\% from the observations (Fig. 7).
We use the $N_{min}$-values derived from the complete spectra, \ie, $N_{min}$ = 20 and 30 for 
NGC 2023 and M17-SW respectively. Noting that $N_{min}$ has a strong effect on the PAH emission 
(Figs. 5$c$ and 6$c$ and Figs. 8$c$ to 11$c$), it is remarkable that the same value allows to match both the absolute 
peak intensity and the profile shape of the 3.3 \mum-band.
The value of $\beta$ is then chosen so as to obtain a good match 
of the width of all bands: we find that $\beta$ = -3.5 is a good compromise. It must be noted, however, that the values 
of $N_{min}$ and $\beta$ are strongly dependent on the adopted profile temperature laws. 

\begin{figure}[!ht]
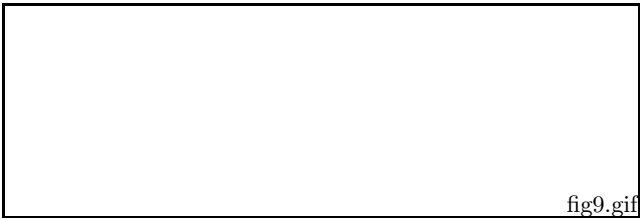

\framebox(240,80)[br]{fig9.gif}
 \caption{ Same as Fig. 8 for the 6.2 \mum-band. In the case of M17-SW, the profile for the 
           $E_{abs}$-approximation has not been represented. 
	  }
\end{figure}

Figs. 8 to 11 show our model representation of the AIB profiles. The contribution of vibrational 
hot bands are not included in our model profiles: in the case of the 6.2 and 11.3 \mum-bands, hot 
bands provide an additional broadening of less than 5 \cm_1 (PJB). 
On Fig. 8$c$, we note the effect of making the $E_{abs}$-approximation (\ie, all molecules are heated by the 
same energy photons, see Sect. 4.3): the lack of the hot tail in the PAH temperature distribution yields a 
narrower band profile. Hence, using this approximation would lead to an underestimate of 
$N_{min}$: \eg, in the case of NGC 2023, $N_{min}=15$ is required to match the observed 3.3 \mum-band if one 
makes the $E_{abs}$-approximation. 
This approximation has, however, negligible consequences for the other bands. 
The effect of a larger $\beta$, -2.25 (corresponding to the classical power-law exponent of -3.5 when the size 
distribution is expressed in terms of the grain radius,
Mathis \etal 1977) is shown as the dashed line on Figs. 8$a$ to 11$a$: a low $\beta$-value is 
clearly required to match the observed band profiles. Decreasing $\beta$, however, has a moderate impact
on the band profiles because the then enhanced contribution of small molecules 
is compensated by a strong increase of the band width leading to a decrease in intensity 
(the integrated cross-sections of the IR bands, $\sigma\times\Delta\nu$, are assumed to be conserved 
with temperature).

 \begin{figure}[!t]
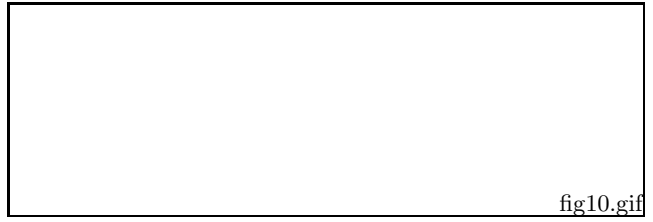

\framebox(240,80)[br]{fig10.gif}
 \caption{ Model results for the 8.6 \mum-band. The data is displayed as in 
           Fig. 8 with the following changes. In $a)$, the solid line shows the model profile if the  
           bandwidth temperature dependence derived from laboratory results is 
           used. The dot-dash line shows the model profile for a hypothetical 
           broadening 5.5 times faster (see text) and the dashed line corresponds 
           to $\beta=-2.25$ case for the fast broadening law. 
           In $b)$, we show the contribution of 
	   small and large PAHs in the case of the fast broadening law.
	   Finally, the display in $c)$ is the same as in $a$ for the M17-SW 
           case except for the dash line where $N_{min}$ = 20 and $\beta=-3.5$ with the 
	   fast broadening law. The unresolved lines at 1035, 1110 and 1245 \cm_1 correspond to
	   0--0 S(3) of H$_2$, a forbidden line of [ArIII] and 0--0 S(2) of H$_2$.
	  }
\end{figure}

The harder radiation field of M17-SW results in PAHs at higher temperatures
(see Fig. 5$d$), consequently the model AIB-profiles are clearly too broad and redshifted 
(see the dashed line in the bottom panel of Figs. 8 to 11) if we use the same size distribution 
as for NGC 2023. For the profile width and position, $N_{min}=30$ is also required in
order to match the observed AIBs of M17-SW. This change of $N_{min}$ has a physical 
interpretation: the photo-dissociation and fragmentation rates rise steeply
with the molecular temperature (L\'eger \etal 1989$a$). The $N_{min}$
values we find are smaller than the theoretical predictions of Allain \etal (1996).

\begin{figure}[!t]
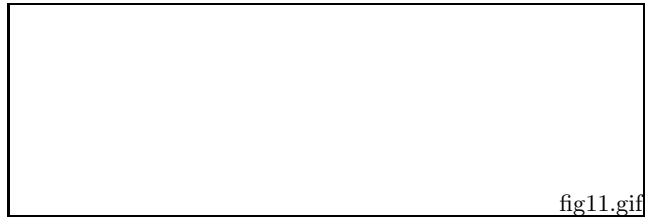

\framebox(240,80)[br]{fig11.gif}
 \caption{ Same as Fig. 9 for the 11.3 \mum-band. Note that the 11 \mum-AIB has not been 
	   withdrawn from the data.
	  }
\end{figure}

Except for the 8.6 \mum-band, all AIB profiles are well explained by the PAH emission model. 
The hot (small) molecules dominate the emission profiles, in particular in the AIB wings; 
they also produce the observed red asymmetry (6.2 and 11.3\mum-bands). The cool (large) molecules
contribute the blue wing of the AIB. We want to stress that this match of the AIBs is obtained
by using the cross-sections and profile temperature dependence measured on small molecules: no 
additional broadening, reflecting the decrease of the level lifetime with increasing 
molecular size (see Sect. 5), has been taken into account here. 

The observed width of the 8.6 \mum-band is not at all reproduced by the temperature broadening 
law derived from laboratory work. In fact, a broadening 5.5 times faster ($\chi_w=6.0\;10^{-2}$ 
\cm_1/K) with temperature would match the observations (see Fig. 10). In this latter case,
we note that {\em (i)} the symmetrical shape of the 8.6 \mum-band is well explained by the slow
redshift of the band centroid with temperature ($\chi_c=0.84\times 10^{-2}$ \cm_1/K whereas the other 
AIBs have 2 to 3$\times 10^{-2}$ \cm_1/K) and {\em (ii)} the emission of large molecules dominates the 
core of the 8.6 \mum-band (conversely to the other AIBs). 

We have not modelled the 7.7 \mum-band because it includes several components whose assignment 
to features in laboratory spectra of PAHs is not straightforward. However, it is interesting 
to note that the observed width (23 to 29 \cm_1) of the 7.6 \mum-band (see Sect. 3) is comparable 
to the width predicted by PJB.

We emphasize here that the model profile of a PAH emission band is the result of the superposition 
of many Lorentz profiles conversely to the implicit assumption behind the decomposition performed in Sect. 3. 
Consider for instance a PAH containing $N_C$ carbon atoms which has been heated 
by a photon of energy $h\nu$: the emerging band profile of this molecule will be the sum 
of all the Lorentz profiles corresponding to the temperatures of the cooling curve, $T_c(t,\nu,N_C)$ 
(\ie, the time integral in Eq. 3). The total band profile of a PAH interstellar population is eventually 
obtained from the weighted sum of the emerging band profiles from all molecules, for all energies of the
exciting photon (see Eqs. 3 and 5). \\

\section{Discussion and Summary}

We have decomposed the ISO-SWS spectra of a number of objects covering a wide range of 
excitation conditions in order to study the individual profiles of the Aromatic 
Infrared Bands (AIBs) between 3 and 13 \mum. All spectra have been decomposed coherently 
into Lorentz profiles and a broadband continuum. We find that the individual 
profiles of the main AIBs at 3.3, 6.2, 8.6 and 11.3 \mum~ are well represented with at most two lorentzians.
The 7.7 \mum-AIB has a more complex shape and requires at least three Lorentz profiles.
We find that the AIB positions and widths are 
stable to within a few \cm_1 (see Table 2) over a range of radiation field hardness
($T_{eff}$=23,000 to 45,000 K) thus confirming results gathered with data at lower spectral 
resolution (Boulanger \etal 1998$a$, Uchida \etal 2000, Chan \etal 2000). 
This spectral decomposition with a small number of Lorentz profiles implicitly assumes that, $(i)$ the AIBs arise 
from a few vibrational bands common to many carriers and, $(ii)$ that most of the bandwidth arises from a 
single carrier. Boulanger \etal (1998$b$) recently proposed that the AIBs are the intrinsic profiles of resonances 
in small carbon clusters containing more than 50 C-atoms. This interpretation can be tested by comparing the AIB profile parameters 
(band position and width) given in this work to laboratory data on relevant species when it becomes available.\\

This spectral decomposition, which separates the AIB emission from the underlying continuum, allowed us to do a detailed comparison 
of the observed AIB spectrum with the predictions of the PAH model where the AIB carriers are free-flying aromatic molecules 
emitting during temperature fluctuations. 
This model uses recent laboratory data and assumes that PAHs are predominantly in cationic form as is expected from 
theoretical work (Bakes \& Tielens 1994, Dartois \& d'Hendecourt 1997).
Within this framework, the position and width of the AIBs are rather explained by a redshift and a broadening of the PAH 
vibrational bands as the temperature of the molecule increases (Joblin \etal 1995).
The observed similarity in the AIB profiles thus requires that some process 
renders the temperature distribution of PAHs rather constant in the interstellar 
regions considered here.
We first derived the temperature distribution
for a population of interstellar PAHs. In particular, we show that the hot tail of the temperature distribution 
of PAHs (which determines the AIB spectrum) depends sensitively on $N_{min}$ and 
$T_{eff}$ which are respectively the size of the smallest PAH (in terms of the number of C-atoms in the 
molecule) and the effective temperature of the exciting radiation field.
The size of the largest PAH, $N_{max}$ (by number of C-atoms), and the index of the power law 
size distribution (expressed in terms of the number of C-atoms per molecule), $\beta$, were found to have little 
impact on the overall AIB emission.

We compared our model results to the data in the two extreme 
cases of our sample: NGC 2023 ($T_{eff}$=23,000 K) and M17-SW ($T_{eff}$=45,000 K). We are able 
to reproduce the spectral distribution of the AIB emission with $N_{min}=20$ in NGC 2023 and $N_{min}=30$ 
in M17-SW and a PAH abundance amounting to 10 and 8\% of the interstellar carbon (these latter values are in good agreement with 
the constraints set by the extinction curve, Joblin \etal 1992, Verstraete \etal 1992, and the infrared emission, D\'esert \etal 
1990 and Dwek \etal 1998). The minimum PAH size $N_{min}$ was found from the requirement that the observed 3.3/11.3 \mum-band 
be well matched.
This change in $N_{min}$ may reflect the enhanced photodestruction rate of small PAHs 
in regions with harder radiation fields (small PAHs reach higher temperatures and evaporate efficiently). 
Using the same $N_{min}$-values, we find that all AIB profile shapes,
except for the 8.6 \mum-band, can be explained with the temperature dependence of the 
band position and width measured in the laboratory. We also show that the PAH size distribution must be very 
steep ($\beta$ = -3.5) in order to account for the observed bandwidths: in our model representation, the AIB profiles
are thus mostly contributed to by small PAH species.
The case of the 8.6 \mum-band is anomalous:
first, the laboratory cross-section is 3 times too weak to explain the observed AIB ratios; 
in addition, we find that the broadening of this band with the temperature should be 5.5 times 
faster in order to match the profile width seen in the SWS data. More laboratory work is required 
to explain this issue. \\

In summary, using the present best knowledge of PAH spectroscopy, it is possible
to account for the AIB spectrum towards bright interstellar objects where the
fraction of singly-ionized PAHs is high. We emphasize that both the spectral shape 
of the AIB spectrum as well as the individual AIB shapes are explained 
consistently with a single set of parameter ($N_{min}$ and $\beta$) values.
In this context, the remarkable stability of the AIB profiles arises if the
hot tail of the PAH temperature distribution remains essentially the same,
whatever the exciting radiation field (the low-temperature side corresponding 
to the larger molecules is practically unaffected by changes in the radiation
field). This requirement is naturally met while considering the photodestruction
of PAHs: when the energy per bond (which is directly proportional to the
molecular temperature) is sufficient the molecule efficiently loses its atoms. 
A consequence of this is that interstellar PAHs would predominantly be destroyed
by thermal evaporation (photo-thermo-dissociation, see L\'eger \etal 1989$a$) rather 
than non-equilibrium processes like direct photodissociation (Buch 1989) or 
Coulomb explosion (Leach 1989). \\

These results must however be placed in the broader context of ISO
observations which include less irradiated regions ($\chi$ = 1 to 1000).
Indeed, ISOCAM-CVF and ISOPHOT-S spectra of faint AIB emission 
highlight the pre-requisites of this work. Namely, two strong assumptions have 
been made in the present model:

\begin{enumerate}
  \item the behaviour of the band profiles with the temperature (measured on small molecules
        only, $N_C\leq$ 50) is the same for all PAH sizes. 
        This assumption applies for all the AIBs except the 3.3 \mum~: in that case the band 
        profile is actually dominated by the contribution of small species (see Fig. 8$b$).
  \item All PAHs have the infrared emissivity of singly ionized cations. 
\end{enumerate}

Assumption 1. is questionable because the intrinsic 
bandwidth ($\Delta\nu_0$, see Sect. 4.4.2) depends on the density of states and on the rate 
of internal conversion processes (\eg, L\'eger \etal 1989$a$).
In particular, at a given internal energy of the molecule, the density of vibrational states is 
known to increase 
steeply for larger and larger molecules (\eg, Cook \& Saykally 1998). Many isoenergetic molecular 
levels may then couple together and eventually lead to a decrease of the level lifetime (Smalley 1983).
Hence, relaxing the first assumption will probably result in a significant 
broadening of the model band profiles; this will require a larger minimum PAH size in order to 
reproduce the observed width of AIBs. Increasing $N_{min}$, the 3.3 \mum-band emission
will decrease rapidly because larger (colder) molecules emit more 
at longer wavelengths and we will then fail to match the intensity of the observed 
3.3 \mum-band. 
This problem will not be alleviated by assuming a PAH ionized fraction of less than 1, 
(\ie, not all PAHs are cations): neutral PAHs, which have strong C-H/C-C band ratios, 
(typically 10 times larger than in cations, Langhoff 1996, Pauzat \etal 1997, Hudgins \&
Sandford 1998), will not reproduce the observed 11.3/7.7 \mum-band ratio. 
This relates to the second assumption of the present model.\\

We have assumed that PAHs are all singly ionized (cations) throughout our
data sample. Such a choice is required to reproduce the observed 
band ratios, in particular the C-H/C-C ratio (Langhoff 1996, Allamandola \etal 1999). 
Yet, this requirement is not fully consistent with theoretical predictions (Dartois \etal 
1997) of the PAH ionized fraction along our lines of sight, namely, 0.5 (50\% of PAHs are 
cations). Furthermore, this problem becomes even more acute in the context of other 
observations spanning a broader range of physical conditions: ISOCAM-CVF and ISOPHOT-S 
data (Boulanger \etal 1998$a$ and 1999; Chan \etal 2000; Miville-Desch\^{e}nes \etal 1999;
Onaka \etal 1999 and 2000, Onaka 2000; Uchida \etal 1998 and 2000) obtained 
towards regions with $\chi$ = 1 to 10$^5$ (from the diffuse interstellar medium to 
\HII~ region interfaces and reflection nebulae) show that the C-H/C-C ratio of the AIBs is 
roughly constant and always corresponds to that of PAH cations. Over such a large range of UV 
radiation flux, the state of PAHs is actually  
expected to change from fully neutral to fully ionized. The ionized fraction of PAHs 
is determined by the value of $\gamma=\chi\,\sqrt{T}/n_e$ with $T$ the gas temperature and $n_e$ 
the electron density (this parameter is proportional to the 
ratio of the ionization rate to the recombination rate of PAHs, Bakes \& Tielens 1994).
To keep the AIB band ratios constant, and hence the ionized fraction of PAHs, requires that 
the ratio $\sqrt{T}/n_e$ can vary over four orders of magnitude in order to compensate the 
variation of $\chi$. In reality, $T$ probably does not change by more than 2 orders of 
magnitudes: then $n_e$, which reflects the density variations mostly, would 
have to vary by three orders of magnitude across the regions observed to keep $\gamma$ constant.
Such large density contrasts at large scale are 
at odds with what is currently known of the structure of these interstellar regions. \\

More generally, the predicted strong variability of the physico-chemical state (ionization, 
dehydrogenation) of PAHs in space is not reflected in the recent ISO spectroscopy database which covers 
a variety of astrophysical conditions. This prediction was based on studies of small 
PAHs ($N_C\leq 50$), the only species currently accessible for investigation either in the 
laboratory or with quantum chemistry. 
As shown by Schutte \etal (1993), the 6 to 16 \mum-AIB spectrum 
is also contributed to by large PAHs ($N_C\geq 100$) while the 3.3 \mum-AIB is 
dominated by the smallest species ($N_C\leq 35$, see Fig. 8): this other solution (large ``PAHs''
or carbon clusters) to the 6 to 16 \mum-AIB spectrum may alleviate the present difficulties. However, the 
spectroscopical and structural properties of carbon clusters are poorly known at present: the ISO 
spectroscopic database is ideally suited to identify plausible candidates and stimulate future work on 
the physics and chemistry of carbon clusters.

{\it Acknowledgments:} We are grateful to Fran\c{c}ois Boulanger, Alain Abergel, 
Christine Joblin and Anthony Jones for many stimulating discussions and our referee for helpful comments. 
We also thank the MPE-SDC
(Garching) and the DIDAC (Groningen) for their constant support in the data reduction 
phase and with the use of SWS-IA3. SWS-IA3 is a joint developement of the SWS consortium. 
Contributing institutes are SRON, MPE, KUL and the ESA Astrophysics division. K. S. gratefully 
acknowledges support from a NATO collaborative Research Grant nr. 951347.

\end{document}